\documentclass[prl, aps, showpacs, reprint]{revtex4-1}

\usepackage[utf8]{inputenc}
\usepackage{graphicx}
\usepackage{amsmath}
\usepackage{amssymb}
\usepackage{braket}
\usepackage{color}
\usepackage{tikz}
\usetikzlibrary{positioning}
\usepackage{pdfpages}

\makeatletter
\AtBeginDocument{\let\LS@rot\@undefined}
\makeatother

\renewcommand\vec[1]{\ensuremath\boldsymbol{#1}}

\newcommand{\Imag}{{\mathrm{Im}}}
\newcommand{\Real}{{\mathrm{Re}}}
\newcommand{\ve}[1]{\boldsymbol{\mathbf{#1}}}

\providecommand{\abs}[1]{\lvert#1\rvert} 

\begin{document}

\title{Spin Pumping and Inverse Spin Hall Voltages from Dynamical Antiferromagnets}

\author{\O yvind Johansen}
\author{Arne Brataas}
\affiliation{Department of Physics, Norwegian University of Science and Technology, NO-7491, Trondheim, Norway}

\date{\today}

\begin{abstract}
Dynamical antiferromagnets pump spins efficiently into adjacent conductors as ferromagnets.  The high antiferromagnetic resonance frequencies represent a challenge for experimental detection, but magnetic fields can reduce these resonance frequencies. We compute the inverse spin Hall voltages resulting from dynamical spin excitations as a function of a magnetic field along the easy axis and the polarization of the driving AC magnetic field perpendicular to the easy axis.  We consider the insulating antiferromagnets MnF$_2$, FeF$_2$, and NiO. Near the spin-flop transition, there is a significant enhancement of the DC spin pumping and inverse spin Hall voltage for the uniaxial antiferromagnets MnF$_2$ and FeF$_2$. In the biaxial NiO, the voltages are much weaker, and there is no spin-flop enhancement of the DC component.
\end{abstract}

\maketitle

Spin pumping is a versatile tool for probing spin dynamics in ferromagnets \cite{Mizukami:jjap2001,Urban:prl2001,Tserkovnyak:prl2002,Brataas:prb2002,Heinrich:prl2003,Tserkovnyak:rmp2005}. The magnitude of the pumped spin currents reveals information about the magnetization dynamics and the electron-magnon coupling at interfaces \cite{Costache:prl2006,Heinrich:prl2011,Kapelrud:prl2013}.  The precessing spins generate a pure spin flow into adjacent conductors. Inside the conductor, the resulting spin accumulation and currents give insight into the spin-orbit coupling. The inverse spin Hall effect (ISHE) is often used to convert the pure spin current into a charge current, which is detected \cite{Mosendz:prl2010,Jiao:prl:2013}. Additionally, the induced non-equilibrium spins can be probed with XMCD measurements \cite{Marcham:prb2013,Baker:prl2016}

Antiferromagnets (AFs)  differ strikingly from ferromagnets \cite{Jungwirth:nnano2016}. There are no stray fields in antiferromagnets, making them more robust against the influence of external magnetic fields. The recent discovery of anisotropic  magnetoresistance \cite{Park:nmat2011,Marti:prl2012,Wang:prl2014}, spin-orbit torques \cite{Zelezny:prl2014}, and electrical switching of an antiferromagnet \cite{Wadley:Science2016} demonstrate the feasibility of antiferromagnets as active spintronics components.

The real benefit of antiferromagnets is that they can enable Terahertz circuits. Unlike ferromagnets,  the resonance frequency of antiferromagnets is also governed by the tremendous exchange energy. We recently demonstrated that the transverse spin conductance controlling spin pumping is as large in antiferromagnet-normal metal junctions (AF$|$N) as in ferromagnet-normal metal junctions \cite{Cheng:prl2014}. Furthermore, this result is valid even when the magnetic system is insulating.  The firm electron-magnon coupling at the interface opens the door for electrical probing of the ultra-fast spin dynamics in antiferromagnets \cite{Cheng:prl2014,Cheng:prl2016}.

Precessing spins in antiferromagnets generate Terahertz currents in adjacent conductors. This ability opens new territory in high-frequency spintronics. Such studies could become influential in gathering vital insight into fast electron dynamics and eventually for a broad range of applications. These electric signals also provide further knowledge about the less explored field of antiferromagnetic spin dynamics. This potential requires thorough exploration; we need to establish several critical aspects. 

 The manner in which spin pumping generates AC and DC inverse spin Hall voltages has yet to be studied in detail. Furthermore, there is a large variety of antiferromagnets and external field configurations that require knowledge beyond the first predictions of the magnitude of the pumped spin current of Ref.\ \onlinecite{Cheng:prl2014}. Recently, researchers explored  spin transport through, e.g.,\ the insulating antiferromagnets NiO and MnF$_2$. Unlike the treatment of Ref.\ \onlinecite{Cheng:prl2014}, in NiO, there are two significant anisotropies to consider. As a starting point in the exploration of high-frequency spintronics, it is also important to tune the resonance frequencies to a lower Gigahertz range for easier detection by conventional electronics. The application of an external magnetic field can lower the resonance frequency. However, the details of the magnetic field and its AC component polarization dependence also remain to be classified, a task that we will perform here.

In this Letter, we compute the inverse spin Hall AC and DC voltages generated by spin pumping. We hope that our studies will further motivate these voltages to be experimentally measured. Such studies will provide a needed deeper insight into antiferromagnetic resonance phenomena, features much less explored than their ferromagnetic counterparts in recent decades. 

\begin{figure}[h!]
\centering
\begin{tikzpicture}
\node[above right] (img) at (0,0) {\includegraphics[width=0.9\linewidth]{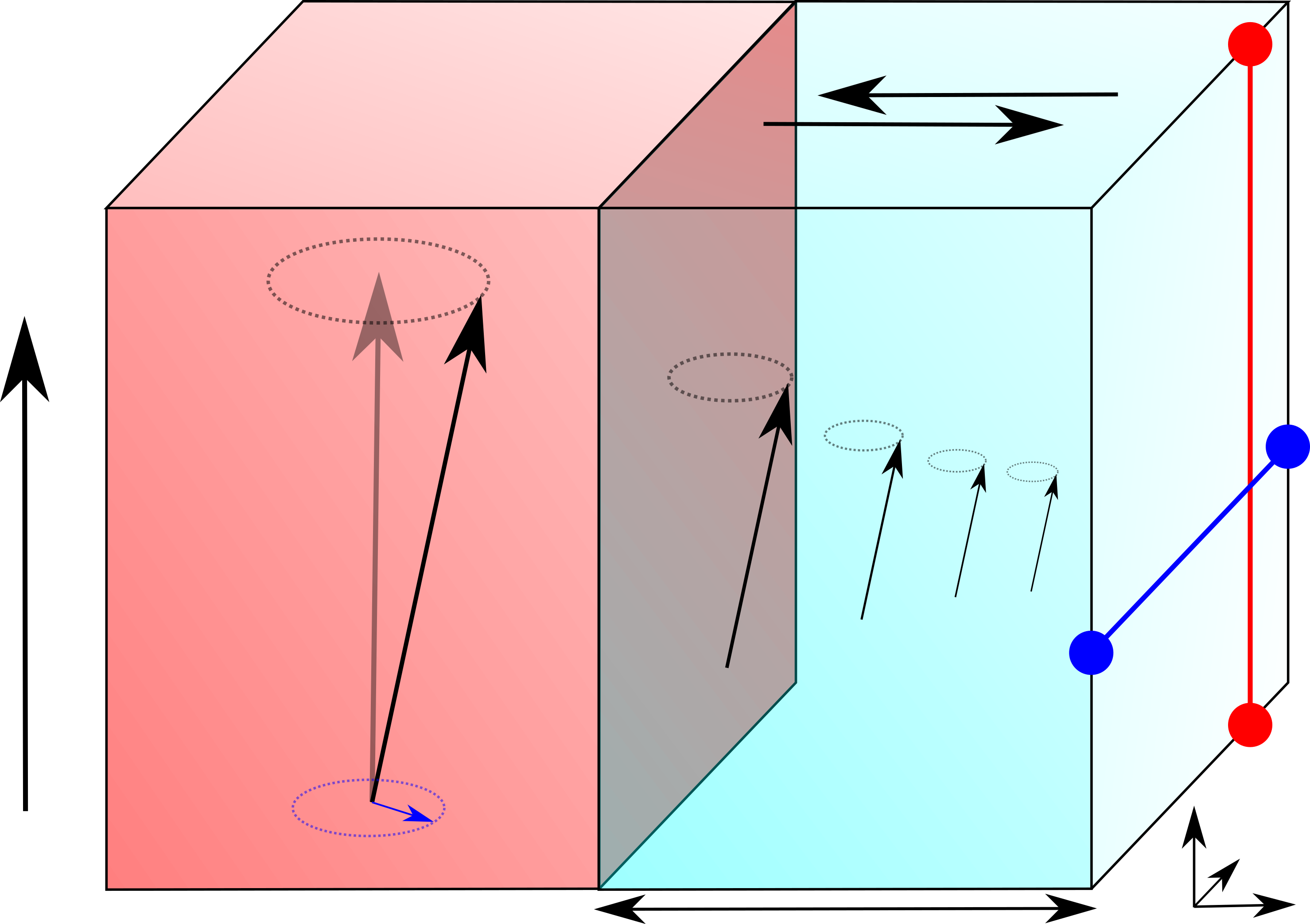}};
\node at (8pt,115pt) {\Large{$\ve{H}_\text{ex}$}};
\node at (155pt,131pt) {\large{$\ve{I}_s^p$}};
\node at (170pt,152pt) {\large{$\ve{I}_s^b$}};
\node at (155pt,35pt) {\Large{$\ve{\mu}_s^N(y,t)$}};
\node at (90pt,32pt) {\textcolor{blue}{\Large{$\ve{m}(t)$}}};
\node at (93pt,85pt) {\Large{$\ve{n}(t)$}};
\node at (58pt,93pt) {\Large{$\ve{n}_0$}};
\node at (200pt,16pt) {$\ve{x}$};
\node at (220pt,-1pt) {$\ve{y}$};
\node at (216pt,15pt) {$\ve{z}$};
\node at (202pt,100pt) {\textcolor{red}{\large{$E_x^\text{AC}$}}};
\node at (202pt,45pt) {\textcolor{blue}{\large{$E_z^\text{DC}$}}};
\node at (148pt,-3pt) {\large{$d_N$}};
\end{tikzpicture}
\caption{The precession of $\ve{m}$ and $\ve{n}$ around their equilibrium values pumps spins into the adjacent normal metal of thickness $d_N$. In turn, the spin accumulation $\ve{\mu}_s^N$ in the normal metal causes a backflow of spins into the antiferromagnet. The spin current in the normal metal causes AC and DC electric fields in the $x$- and $z$-directions respectively through the inverse spin Hall effect.}
\label{fig:Bilayer}
\end{figure}
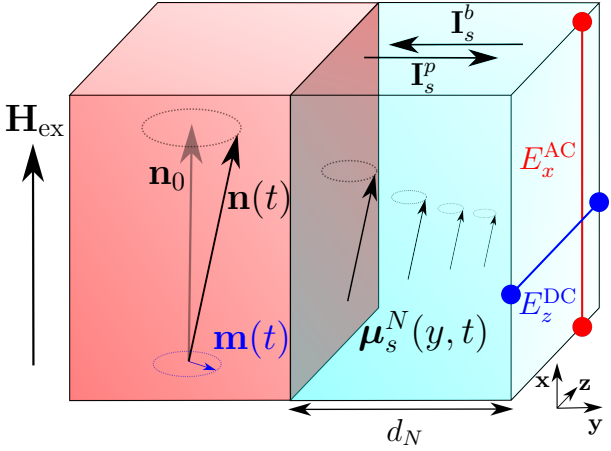

We consider an insulating antiferromagnet-normal metal bi-layer, as illustrated in Fig. \ref{fig:Bilayer}. We also consider a variety of magnetic anisotropies and magnetic field configurations and strengths.  Therefore, the results apply to more complex systems such as biaxial antiferromagnets with elliptical precessional modes.  The model also accounts for spin backflow due to the spin accumulation in the metal. We also study how the inverse spin Hall voltages depend on the polarization of the AC magnetic field for different systems, which we find to have a strong influence on the resulting signal. Our main findings are that, when applying an external magnetic field along the easy axis close to the spin-flop transition, we can decrease the resonance frequency while simultaneously significantly increasing the inverse spin Hall signal. The increase in the signal can even overcome the previously anticipated limiting factor in antiferromagnet spin pumping: the ratio of the anisotropic energy to the exchange energy \cite{Cheng:prl2014}.

We consider a small antiferromagnet in the macrospin limit whereby all spin excitations are homogeneous. The antiferromagnet has two sublattices, with temporal magnetizations ${\bf M}_1$ and ${\bf M}_2$. The dynamics are described by the staggered magnetizations ${\bf L}=\left({\bf M}_1-{\bf M}_2\right)/2=L {\bf n}$ and the magnetization ${\bf M}=\left({\bf M}_1+{\bf M}_2\right)/2=L {\bf m}$. These fields satisfy the constraints ${\bf n}^2 + {\bf m}^2=1$ and ${\bf n} \cdot {\bf m}=0$. At equilibrium,  the sublattice magnetizations are anti-parallel. An AC magnetic field, with a general polarization, drives the magnetic moments at resonance.

The antiferromagnets that we consider are described by the free energy 
%
%
\begin{align}
\nonumber F = &\frac{L V}{\gamma} \big[ \omega_E \left(\bf{m}^2 - \bf{n}^2 \right) +  \omega_{\perp} \left(m_z^2 + n_z^2 \right) - \omega_{\parallel} \left(m_x^2 + n_x^2 \right)  \\
&-  2 \omega_{x} m_x - 2\omega_{y} m_y - 2\omega_{z} m_z \big]  \,  ,
\label{eq:FreeEnergy}
\end{align}
%
%
where  $\gamma$ is the gyromagnetic ratio, $V$ is the volume of the antiferromagnet, $\omega_E \ge 0 $ is the exchange frequency, and $\omega_\perp \ge 0$ and $\omega_\parallel \ge 0 $ are the hard axis ($z$-axis) and easy axis ($x$-axis) anisotropy frequencies. The frequency $\omega_{x}$ quantifies the influence of the external magnetic field along the easy axis, whereas $\omega_{y}$ and $\omega_{z}$ quantify the influence of the AC magnetic field in the $yz$-plane. In Table \ref{tab:frequencies}, we list the exchange and anisotropy frequencies for MnF$_2$, FeF$_2$ and NiO. 
\begin{table}[h!]
	\centering
		\caption{Exchange and anisotropy frequencies.} 
	\begin{tabular}{l l l l} \hline
	Material & $\omega_E$ ($10^{12}$ s$^{-1}$)& $\omega_{\parallel}$ ($ 10^{12}$ s$^{-1}$) & $\omega_{\perp}$ ($10^{12}$ s$^{-1}$)\\ \hline
	MnF$_2$ \cite{Ross:TUMunchen:2013} & 9.3 & 1.5 $\cdot 10^{-1}$ & - \\
 	FeF$_2$ \cite{Ohlmann:prb:1961} & 9.5 & 3.5 & - \\
 	NiO \cite{Satoh:prl:2010,Hutchings:prb:1972} & 1.7$\cdot 10^{2}$& 2.3 $\cdot 10^{-3}$ & 1.3 $\cdot 10^{-1}$ \\ \hline
	\end{tabular}
	\label{tab:frequencies}
\end{table}

The dynamic Landau-Lifshitz-Gilbert equations that describe the precession of $\ve{n}$ and $\ve{m}$ are
\begin{subequations}
\label{eq:LLGnm}
\begin{align}
\dot{{\bf n}}  & = \frac{1}{2}\left(\boldsymbol{\omega}_m \times {\bf n} + \boldsymbol{\omega}_n \times {\bf m}\right) + \boldsymbol{\tau}_n \, , \\
\dot{{\bf m}} & = \frac{1}{2}\left(\boldsymbol{\omega}_n \times {\bf n} + \boldsymbol{\omega}_m \times {\bf  m}\right) + \boldsymbol{\tau}_m \, ,
\end{align}
\end{subequations}
with the effective fields $\boldsymbol{\omega}_n = -(\gamma/L) \partial F/\partial {\bf n}$ and $\boldsymbol{\omega}_m = - (\gamma/L) \partial F / \partial {\bf m}$. The dissipation and spin-pumping torques are 
\begin{subequations}
\begin{align}
\boldsymbol{\tau}_n =  \alpha \left[ {\bf n}  \times \dot{{\bf m}}  + {\bf m} \times  \dot{{\bf n}}  \right] \, , \\
\boldsymbol{\tau}_m = \alpha \left[ {\bf n} \times \dot{{\bf n}}  + {\bf m} \times \dot{{\bf m}}  \right] \, , 
\end{align}
\end{subequations}
where the total Gilbert damping coefficient $\alpha$ is a sum of the intrinsic damping and the spin-pumping-enhanced damping: $\alpha= \alpha_0 + \alpha_\text{SP}$ \cite{Hals:prl2011,Cheng:prl2014}.

A linear response expansion around the equilibrium values of $\ve{n}$ and $\ve{m}$ determines the antiferromagnetic resonance (AFMR) frequencies. 
For simplicity, we only present the resonance frequencies in the exchange limit $\omega_\parallel,\omega_\perp\ll\omega_E$. This limit is valid for many antiferromagnets but not for FeF$_2$ due to a large anisotropy. In our numerical calculations below, we do not make this approximation.
In the exchange limit, the four resonance frequencies below spin-flop are \cite{Yosida:1952}
\begin{align}
\omega_\text{res}^2 \approx \omega_x^2+\omega_0^2
\pm \sqrt{\omega_E^2\omega_\perp^2+4\omega_x^2\omega_0^2} \, ,
\end{align}
where $\omega_0^2 = \omega_E\left(2\omega_\parallel+\omega_\perp\right)$. The critical field strength at which the spin-flop transition occurs is $\abs{\omega_{x}^\text{crit}}= \sqrt{\omega_\parallel(2\omega_E+\omega_\parallel)}$ in both uniaxial and biaxial antiferromagnets. We will only consider magnetic fields below this value. 

Herein, we focus on the right-handed low-energy mode since we want to decrease the resonance frequency. In the absence of an external magnetic field, the resonance frequency of this mode is 0.27 THz for MnF$_2$, 1.41 THz for FeF$_2$, and 0.14 THz for NiO. By applying a magnetic field close to the spin-flop transition, we can reduce these resonance frequencies down to the GHz range. Such a reduction should enable detection of AFMR and the resulting significant spin-pumping-induced AC and DC ISHE voltages.

The pumped spin current from a dynamical antiferromagnet into a normal metal is \citep{Cheng:prl2014}
\begin{equation}
\ve{I}_s^p= \frac{\hbar g_\perp}{2\pi}\left(\ve{n}\times\dot{\ve{n}}+\ve{m}\times\dot{\ve{m}}\right) \, ,
\label{eq:spinpumping}
\end{equation}
where $g_\perp$ is the transverse ("mixing") conductance.  The spin pumping from the antiferromagnetic insulator causes a spin accumulation in the normal metal, which in turn produces a spin backflow current \cite{Jiao:prl:2013}. 
In antiferromagnetic insulators, the backflow spin currents within the sublattices add constructively \cite{Gomonay:prb2010,Cheng:prl2014}:
\begin{align}
\ve{I}_s^b = -\frac{g_\perp}{2\pi}\left(\ve{m}\times (\ve{\mu}_s^N\times\ve{m})+\ve{n}\times (\ve{\mu}_s^N\times\ve{n})\right) \, , 
\label{eq:spinbackflow}
\end{align}
where $\ve{\mu}_s^N$ is the spin accumulation in the normal metal.

The most significant contributions to the spin current are second order in the deviations from equilibrium along the easy axis and first order along the perpendicular directions. Nevertheless, the leading-order terms in the total spin current only depend on the first-order deviations of the magnetic moments from their equilibrium values, $\ve{n}_0=\ve{e}_x$ and $\ve{m}_0=\ve{0}$. It is therefore sufficient to consider the linear response expansions
\begin{subequations}
\label{eq:linear_response}
\begin{align}
	\label{eq:n}
	\ve{n} &= \ve{n}_0 + \frac{1}{2}\left(\delta\ve{n} e^{i\omega t}+\delta\ve{n}^*e^{-i\omega t}\right) \, , \\
	\label{eq:m}
	\ve{m} &= \frac{1}{2}\left(\delta\ve{m} e^{i\omega t}+\delta\ve{m}^*e^{-i\omega t}\right) \, ,
\end{align}
\end{subequations}
where the transverse deviations are $\delta\ve{n}=\delta n_y\ve{e}_y+\delta n_z\ve{e}_z$ and $\delta\ve{m}=\delta m_y\ve{e}_y+\delta m_z\ve{e}_z$. $\omega$ is the driving frequency of the AC magnetic field. Consequently, to leading order, we can disregard the dependence of the spin backflow on $\ve{m}$.

The spin accumulation $\ve{\mu}_s^N$ is a solution of the spin diffusion equation
\begin{equation}
\label{eq:spindiffusion}
\frac{\partial\ve{\mu}_s^N(\ve{r},t)}{\partial t}=\gamma_N\ve{H}_{\text{ex}}\times\ve{\mu}_s^N+D_N\frac{\partial^2\ve{\mu}_s^N}{\partial y^2}-\frac{\ve{\mu}_s^N}{\tau_{\text{sf}}^N} \, ,
\end{equation}
where the terms on the right-hand side of Eq.\ \eqref{eq:spindiffusion} are properties of the normal metal such as the diffusion coefficient $D_N$, the gyromagnetic ratio $\gamma_N$, and the spin-flip relaxation time $\tau_{\text{sf}}^N$, and $\ve{H}_{\text{ex}}$ is the external magnetic field.
The boundary conditions for $\ve{\mu}_s^N$ require that the spin current vanishes at the outer edge of the normal metal ($y=d_N$) and that the current is continuous across the antiferromagnet-normal metal interface ($y=0$). 
The diffusion equation can be solved in position-frequency space \cite{Jiao:prl:2013,Johnson:prb:1988} in terms of the Fourier components of the total spin current $\ve{I}_s^N=\ve{I}_s^p+\ve{I}_s^b$ at $y=0$.

The spin current in the normal metal causes a charge current perpendicular to the spin current's direction and polarization through the ISHE. This charge current is given by \cite{Saitoh:apl:2006,Mosendz:prb:2010}
\begin{equation}
\ve{j}_c^{\text{ISHE}}(y,t)=\theta_N \frac{2e}{A\hbar}\ve{e}_y\times\ve{I}_s^N(y,t) \, ,
\end{equation}
where $\theta_N$ is the spin Hall angle in the normal metal and $A$ is the area of the AF$|$N interface. Since the system is an open circuit, the charge current accumulates charges at the interfaces. In turn, a generated electric field ensures that the net charge current through the metal vanishes. To determine this electric field, we integrate the charge current $\ve{j}_c^{\text{ISHE}}$ over the metallic system to find the electric field needed to cancel the charge current. See the Supplementary Material \cite{Supplementary} for the full derivation. The DC component of this electric field becomes
\begin{equation}
E_z^{\text{DC}} = \varepsilon_N\left(1-\frac{1}{\cosh{\left(d_N/\lambda_{\text{sd}}^N\right)}}\right)\mu_0^x \, ,
\end{equation}
the first harmonic AC component is
\begin{align}
\nonumber E_x^{\text{AC}}(t) &=  \varepsilon_N\Real \biggl[\biggl(\frac{\mu_1^z+i\mu_1^y}{\cosh{\left(\kappa_3(\omega)d_N\right)}}\\
&+\frac{\mu_1^z-i\mu_1^y}{\cosh{\left(\kappa_2(\omega)d_N\right)}} - 2\mu_1^z \biggr)e^{i\omega t}\biggr] \, , 
\end{align}
and the second harmonic AC component is
\begin{equation}
E_z^{\text{AC}}(t) = 2\varepsilon_N \Real \left[\left(1-\frac{1}{\cosh{\left(\kappa_1(2\omega)d_N\right)}}\right)\mu_2^x e^{2i\omega t}\right] \, .
\end{equation}
Here, we have introduced the conversion coefficient $\varepsilon_N=\theta_N e \nu D_N/(\sigma_N d_N)$, where $\sigma_N$ is the conductivity of the normal metal.
The factors $\mu_n^{x/y/z}$ are the $n$-th Fourier components of the spin accumulation at the AF$|$N interface ($y=0$). We compute that they are
\begin{subequations}
\begin{align}
\mu_{1}^y &= -\frac{i\hbar\omega g_\perp}{4\pi}\frac{\left(\Gamma_2\left(\omega\right)+\frac{g_\perp}{2\pi}\right)\delta n_z+\Gamma_3\left(\omega\right)\delta n_y}{\left(\Gamma_2\left(\omega\right)+\frac{ g_\perp}{2\pi}\right)^2+\Gamma_3^2\left(\omega\right)} \, , \\
\mu_{1}^z &= \frac{i\hbar\omega g_\perp}{4\pi}\frac{\left(\Gamma_2\left(\omega\right)+\frac{g_\perp}{2\pi}\right)\delta n_y-\Gamma_3\left(\omega\right)\delta n_z}{\left(\Gamma_2\left(\omega\right)+\frac{ g_\perp}{2\pi}\right)^2+\Gamma_3^2\left(\omega\right)} \, , \\
\mu_2^x &= \frac{g_\perp}{4\pi\Gamma_1(2\omega)}\left(\mu_1^y\delta n_y+\mu_1^z\delta n_z\right) \, ,
\end{align}
\label{eq:spinaccAC}
\end{subequations}
for the first and second harmonic AC components, and
\begin{align}
\nonumber\mu_0^x = &\frac{g_\perp}{2\pi\Gamma_1(0)}\bigl[\Real\left(\mu_1^y\delta n_y^*+\mu_1^z\delta n_z^*\right) \\
&-\hbar\omega\Imag\left(\delta n_y^*\delta n_z+\delta m_y^*\delta m_z\right)\bigr] \, ,
\label{eq:spinaccDC}
\end{align}
for the DC component. 
All other components of the spin accumulation vanish. The components of the spin accumulation of Eqs.\ \eqref{eq:spinaccAC} and \eqref{eq:spinaccDC} are expressed in terms of the functions
\begin{subequations}
\begin{align}
\Gamma_1(\omega) &= \frac{1}{2}\hbar\nu A D_N\Lambda_1(\omega),\\
\Gamma_2(\omega) &= \frac{1}{4}\hbar\nu A D_N\left[\Lambda_2(\omega)+\Lambda_3(\omega)\right] \, , \\
\Gamma_3(\omega) &= \frac{i}{4}\hbar\nu A D_N\left[\Lambda_2(\omega)-\Lambda_3(\omega)\right] \, ,
\end{align}
\end{subequations}
with $\Lambda_i(\omega)=\kappa_i(\omega)\tanh\left[\kappa_i(\omega)d_N\right]$.
Here, we have defined $\kappa_1^2 = \left(1+i\omega\tau_{\text{sf}}^N\right)/\left(\lambda_{\text{sd}}^N\right)^2$, $\kappa_{(2,3)}^2 = \kappa_1^2\mp i \gamma_N H_{\text{ex}}/ D_N$, the spin diffusion length $\lambda_{\text{sd}}^N = \sqrt{D_N \tau_{\text{sf}}^N}$ and the one-spin density of state $\nu$.
Note that $\mu_2^x$ and consequently $E_z^\text{AC}$ vanish in the absence of a magnetic field ($\Gamma_3(\omega)=0$) and when the precession of the staggered magnetization is circular ($\delta n_z = \pm i\delta n_y$).

We will now use our model to compute the ISHE signal as a function of external magnetic fields in an AF$|$Pt bilayer. By inserting the linear response ansatz of Eq.\ \eqref{eq:linear_response} into the LLG equations in Eq.\ \eqref{eq:LLGnm}, we determine the functions $\delta\vec{n}$ and $\delta\vec{m}$. The components of the AC magnetic field that drives these perturbations are given by $\omega_j = \abs{\omega_j} \exp (i\omega t + i\theta_j)$ for $j=y,z$. The phase difference $\theta_z-\theta_y$ determines the polarization of the AC field, and significantly affects the resulting spin current. In our calculations, we let $\abs{\omega_y}=\abs{\omega_z}$.

As the material properties of Pt, we use $\tau_\text{sf}^N=0.01$\ ps \cite{Jiao:prl:2013}, $\nu =4.55\cdot 10^{47} \text{ J}^{-1}$\ $\text{m}^{-3}$ \cite{papaconstantopoulos1986handbook}, $\sigma_N=5\cdot 10^6$\ $(\mathrm{\Omega m})^{-1}$ \cite{Liu2011}, $\lambda_\text{sd}^N = 1.5$\ nm, and $\theta_N=0.075$ \cite{Meyer:apl:2014}. These properties are at 10\ K. The transverse conductance $g_\perp$ has yet to be determined experimentally for antiferromagnets. However, it should be of the same order of magnitude as that of a ferromagnetic or ferrimagnetic material \cite{Cheng:prl2014}. A reasonable estimate of this parameter is therefore $g_\perp/A = 3\cdot 10^{18}$\ m$^{-2}$ \cite{Yoshino2011,Haertinger:prb:2015}, which we use in the following.  Experimental measurements of $g_\perp$ are needed and are further motivated by the present calculations.

The magnitude of the ISHE signal depends on the thickness of the Pt layer. It increases approximately linearly with $d_N$ for $d_N/\lambda_\text{sd}^N\ll 1$ and is inversely proportional to $d_N$ for $d_N/\lambda_\text{sd}^N\gg 1$. This qualitative behavior is similar to that in ferromagnetic/normal metal bilayers, cf. Fig. 3(a) in Ref. \onlinecite{Jiao:prl:2013}. The peak of the ISHE signal is at some value $d_N \sim \lambda_\text{sd}^N$, and for our choice of parameters, it peaks at $d_N\approx 0.8 \lambda_\text{sd}^N = 1.2$ nm. We use this thickness of the Pt layer for the remaining calculations. The optimal thickness $d_N$ weakly depends on the value of $g_\perp/A$ and should therefore be determined experimentally.

\begin{figure}[h!]
\centering
\begin{tikzpicture}
\node[above right] (img) at (0,0) {\includegraphics[width=0.98\linewidth]{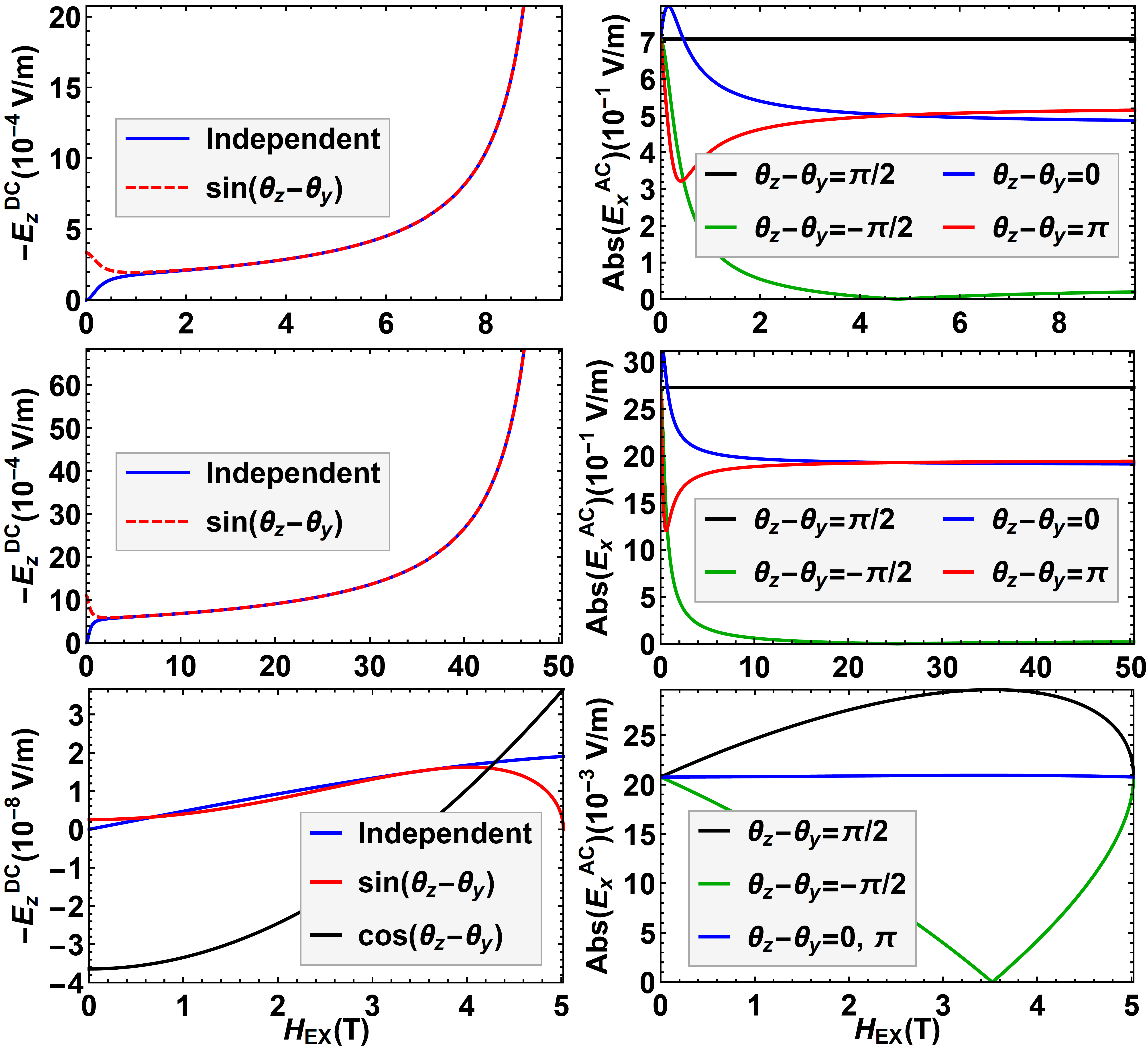}};
\node at (32pt,208pt) {$\ve{(a)}$};
\node at (230pt,208pt) {$\ve{(b)}$};
\node at (32pt,138pt) {$\ve{(c)}$};
\node at (230pt,136pt) {$\ve{(d)}$};
\node at (32pt,69pt) {$\ve{(e)}$};
\node at (230pt,69pt) {$\ve{(f)}$};
\end{tikzpicture}
\caption{DC and first harmonic AC components of the ISHE electric field for MnF$_2$ [(a) and (b)], FeF$_2$ [(c) and (d)], and NiO [(e) and (f)] as a function of external magnetic field along the easy axis for different polarizations of the AC magnetic field. The AC field is 1 mT, and $\alpha=0.01$.}
\label{fig:ISHE_Uniaxial}
\end{figure}
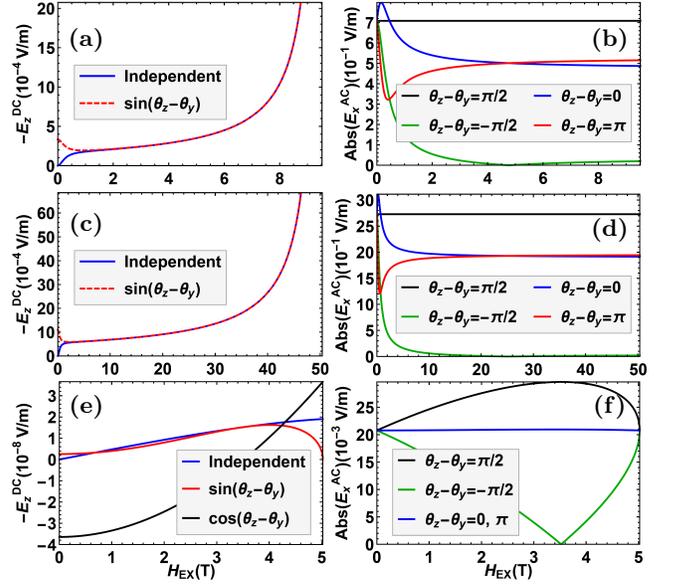

Fig.\ \ref{fig:ISHE_Uniaxial} plots the DC and the first harmonic AC components of the ISHE electric field as a function of the magnetic field. In the uniaxial antiferromagnets, MnF$_2$ and FeF$_2$, one contribution to the DC signal is independent of the AC magnetic field polarization, and the other contribution is proportional to $\sin (\theta_z-\theta_y)$. At high magnetic fields, these contributions are equal in magnitude but add constructively or destructively, depending on the circular polarization of the AC magnetic field.

Ref. \onlinecite{Cheng:prl2014} demonstrated that the pumped DC spin current in uniaxial antiferromagnets at resonance is suppressed by the factor $\sqrt{\omega_\parallel/\omega_E}$. Since $\sqrt{\omega_\parallel/\omega_E}$ is significantly larger in FeF$_2$ (0.61) than in MnF$_2$ (0.13), it was believed that FeF$_2$ gives a stronger signal than does MnF$_2$. However, with our present additional insight, we reach the opposite conclusion at finite magnetic fields. We find that when $\omega_x\rightarrow \omega_x^\text{crit}$, the DC signal diverges as $(\omega_x^\text{crit} - \omega_x)^{-1}$. The utilization of the divergence is a better route toward enhancing the ISHE signal than increasing $\sqrt{\omega_\parallel/\omega_E}$. This implies that MnF$_2$ is a more promising candidate than FeF$_2$ because the spin-flop field of MnF$_2$ (9.5 T) is easier to achieve experimentally than is that of FeF$_2$ (50.4 T). 

Unlike the DC component, the first harmonic AC component is independent of the AC magnetic field polarization in the absence of a uniform external magnetic field and converges toward a finite value as $\omega_x\rightarrow \omega_x^\text{crit}$. The signal when the polarization is circular ($\theta_z-\theta_y=\pi/2$) gives the largest DC signal (and AC signal for sufficiently large magnetic fields). Furthermore, this curve becomes independent of the magnetic field. The origin of this is a complicated compensation between the diverging contributions from the out-of-equilibrium fields and the vanishing resonance frequency around the spin-flop transition.

In NiO, the dominant AC magnetic field contribution is linear in the polarization, which is proportional to $\cos (\theta_z-\theta_y)$. Such a feature appears when there is a hard axis, and the precession is in the easy plane. The linear contribution dominates when $\omega_\perp/\sqrt{\omega_E\omega_\parallel}\gg\alpha$. In biaxial antiferromagnets, we find that the pumped current is governed by the scaling factor $\alpha\omega_\parallel/\omega_\perp$ instead of $\sqrt{\omega_\parallel/\omega_E}$. In discussing the strength of the spin-pumping signals, we should also note that, in both uniaxial and easy-plane antiferromagnets, the signal is inversely proportional to $\omega_E$. Since $\omega_\parallel/\omega_\perp\sim 0.02$ and since $\omega_E$ is exceptionally large in NiO, the DC spin pumping signal is weak in comparison to that of MnF$_2$ and FeF$_2$.
In addition, the DC signal in NiO does not exhibit a divergence as $\omega_x\rightarrow \omega_x^\text{crit}$. 
The first harmonic AC component in NiO is independent of the polarization of the AC magnetic field at the spin-flop transition, unlike the uniaxial antiferromagnets, but the magnitude is small.
We do not present the second harmonic AC voltage since it is minimal (and in many cases identically zero) compared to the other voltages. The exception is for NiO just around spin flop, where it can be the same order of magnitude as the DC voltage; however, this is still a very weak signal. Our results imply that uniaxial antiferromagnets are preferred candidates for the observation of spin pumping compared to hard-axis antiferromagnets such as NiO.

Ref. \onlinecite{Ross:TUMunchen:2013} conducted preliminary spin-pumping experiments for a MnF$_2|$Pt system. However,  they attributed the dominant DC signal to microwave rectification and not spin pumping.  Nevertheless, they observed a small change in the signal upon reversal of the magnetic field, which is consistent with spin pumping. 

We propose a different experimental geometry to enhance the spin-pumping signal. The use of the AC magnetic field in a plane perpendicular to the easy axis and a polarization $\theta_z-\theta_y=\pi/2$ increases the DC ISHE signal by a factor of 4. Additionally, by reducing the thickness of the Pt layer from 7 nm to the thickness where the ISHE signal attains its maximum (in our calculations, this is 1.2 nm), we can further amplify the signal by a factor of 2. Together, these improvements will increase the signal strength by an order of magnitude. Whether the signal is due to spin pumping can then easily be tested by the dependence on the polarization of the AC magnetic field according to our model. A circular polarization with $\theta_z-\theta_y=\pi/2$ doubles the signal strength compared to a linear polarization. On the other hand, a circular polarization with $\theta_z-\theta_y=-\pi/2$ results in no DC spin pumping. In contrast, microwave rectification effects should be much less sensitive to the polarization.

In summary, we computed the inverse spin Hall signal as a result of spin pumping and spin backflow in an AF$|$N bi-layer. Our results apply to any polarization of the AC magnetic field and precessional motion of the magnetizations, and the results can also be used in more complex biaxial antiferromagnets. We demonstrate that the DC signal increases substantially near the spin-flop transition in uniaxial antiferromagnets. Furthermore, the signal strongly depends on the polarization of the AC magnetic field. We also suggest an improved experimental geometry that considerably enhances the DC signal resulting from spin pumping.

The research leading to these results has received funding from the European Research Council via Advanced Grant number 669442 ``Insulatronics'', EU FET ``Transpire'' via grant no. 737038 and The Research Council of Norway via grant no. 239926/F20.

\appendix

\bibliography{bibliography}{}

\cleardoublepage


\section*{Supplementary material}

# Supplementary Material for "Spin Pumping and Inverse Spin Hall Voltages from Dynamical Antiferromagnets"

Øyvind Johansen and Arne Brataas

*Department of Physics, Norwegian University of*

*Science and Technology, NO-7491, Trondheim, Norway*

(Dated: February 13, 2017)

## Abstract

Precessing spins can induce currents between antiferromagnetic insulators and normal metals. We compute these AC and DC spin currents across the interface, including the effects of spin backflow. We also calculate the resulting AC and DC inverse spin Hall electric fields in the normal metal. We express all currents and fields in terms of the dynamics of the staggered field and the magnetization.

**TOTAL SPIN CURRENT AND INVERSE SPIN HALL VOLTAGES**

Spin-pumping is the emission of spin-currents into adjacent conductors from precessing spins. Ref. 1 introduced a quantitative theory for the resulting spin-current through ferromagnet-normal metal interfaces. The spin flow causes a spin accumulation in the normal metal. Ref. 2 computed the resulting DC spin accumulation. There are also AC components and Ref. 3 studied these in detail, and furthermore, included the inverse spin Hall effect in the normal metal. We will generalize this approach by replacing the ferromagnets with antiferromagnetic insulators.

We disregard the imaginary part of the transverse ("mixing") conductance since it is likely to be small in most circumstances[4]. Then, the pumped spin-current is[4]

$$\mathbf{I}_s^p = \frac{\hbar g_\perp}{2\pi} \left( \mathbf{n} \times \dot{\mathbf{n}} + \mathbf{m} \times \dot{\mathbf{m}} \right), \tag{1}$$

where $g_\perp$ is the transverse conductance, $\mathbf{n} = (\mathbf{m}_1 - \mathbf{m}_2)/2$ is the staggered field and $\mathbf{m} = (\mathbf{m}_1 + \mathbf{m}_2)/2$ is the magnetization. At equilibrium, the staggered field is along the easy axis, $\mathbf{n}_0 = \mathbf{e}_x$, and the magnetization vanishes.

Next, we need to take into account the backflow spin current into the antiferromagnet resulting from the spin accumulation. In antiferromagnets, the backflow spin current is similar to the case of ferromagnets since the spin currents from the two sub-lattices add constructively

$$\mathbf{I}_s^b = -\frac{g_\perp}{4\pi} \left[ \mathbf{m}_1 \times \left( \boldsymbol{\mu}_s^N \times \mathbf{m}_1 \right) + \mathbf{m}_2 \times \left( \boldsymbol{\mu}_s^N \times \mathbf{m}_2 \right) \right], \tag{2a}$$

$$= \frac{g_\perp}{2\pi} \left[ \mathbf{m} \left( \boldsymbol{\mu}_s^N \cdot \mathbf{m} \right) + \mathbf{n} \left( \boldsymbol{\mu}_s^N \cdot \mathbf{n} \right) - \boldsymbol{\mu}_s^N \right], \tag{2b}$$

where $\boldsymbol{\mu}_s^N$ is the spin accumulation in the normal metal. The backflow spin current of Eq. (2b) and the pumped spin current of Eq. (1) are related by Onsager reciprocity relations[4].

To compute the spin backflow we need to determine the spatiotemporal variation of the spin accumulation in the normal metal. This spin accumulation should fulfil the spin diffusion equation

$$\frac{\partial \boldsymbol{\mu}_s^N(\mathbf{r}, t)}{\partial t} = \gamma_N \mathbf{H}_{\text{ex}} \times \boldsymbol{\mu}_s^N + D_N \frac{\partial^2 \boldsymbol{\mu}_s^N}{\partial y^2} - \frac{\boldsymbol{\mu}_s^N}{\tau_{\text{sf}}^N}, \tag{3}$$

where the terms on the right hand side of Eq. (3) are properties of the normal metal such as the diffusion coefficient $D_N$, the gyromagnetic ratio $\gamma_N$, the spin-flip relaxation time $\tau_{\text{sf}}^N$, and the external magnetic field $\mathbf{H}_{\text{ex}} = \mathbf{e}_x \omega_x / \gamma_{AF}$. We have disregarded the small off-resonance

AC component of the magnetic field. As a consequence, the magnetic field only causes spin precession that couples the transverse, $y$- and $z$-components of the spin accumulation.

The spin diffusion equation of Eq. (3) needs to be supplemented by boundary conditions. At the outer edge of the normal metal, $y = d_N$, the spin current vanishes. In contrast, the spin current is continuous across the antiferromagnet-normal metal interface, where $y = 0$. We carry out a Fourier transformation of the spin accumulation in time. With the two boundary conditions, the solution to Eq. (3) is[3,5]:

$$\boldsymbol{\mu}_s^N(y,\omega) = \sum_{i=1}^{3} \mathbf{e}_i \frac{\cosh\left[\kappa_i(y - d_N)\right]}{\sinh\left[\kappa_i d_N\right]} \frac{2 j_{is}(y=0,\omega)}{\hbar \nu D_N \kappa_i}. \tag{4}$$

In Eq. (4), we use a circular basis, $\mathbf{e}_1 = \mathbf{e}_x$, $\mathbf{e}_2 = \mathbf{e}_- = (\mathbf{e}_y - i\mathbf{e}_z)/\sqrt{2}$, $\mathbf{e}_3 = \mathbf{e}_+ = (\mathbf{e}_y + i\mathbf{e}_z)/\sqrt{2}$. In this basis, we introduce the spin current density components at the interface: $j_{1s} = I_s^x/A$, $j_{2s} = (I_s^y + iI_s^z)/(\sqrt{2}A)$ and $j_{3s} = (I_s^y - iI_s^z)/(\sqrt{2}A)$, where $A$ is the interface cross section and $I_s^{x/y/z} = \left[\mathbf{I}_s^p(y=0,\omega) + \mathbf{I}_s^b(y=0,\omega)\right] \cdot \mathbf{e}_{x/y/z}$ are the cartesian Fourier components of the total spin current at the interface. We have also defined the quantities $\kappa_1^2 = \left(1 + i\omega \tau_{sf}^N\right)/\left(\lambda_{sd}^N\right)^2$, $\kappa_{(2,3)}^2 = \kappa_1^2 \mp i\gamma_N \omega_x/(\gamma_{AF} D_N)$ and the spin diffusion length $\lambda_{sd}^N = \sqrt{D_N \tau_{sf}^N}$.

Temporal variations of the staggered field and the magnetization drive the spin currents. We characterize the out-of-equilibrium deviations of these fields by a perturbation parameter $\delta$. We will now consider how the different components scale with the small variations proportional to $\delta$ during the spin dynamics. Since $\mathbf{m}_1$ and $\mathbf{m}_2$ are real unit vectors, $\mathbf{n}$ and $\mathbf{m}$ must fulfil $\mathbf{n}^2 + \mathbf{m}^2 = 1$ and $\mathbf{n} \cdot \mathbf{m} = 0$. These conditions are satisfied to second order in the perturbation parameter $\delta$ when we expand the fields as

$$\mathbf{n} = \left(1 - \delta^2 n_x^{(2)}(t)\right) \mathbf{e}_x + \left(\delta n_y^{(1)}(t) + \delta^2 n_y^{(2)}(t)\right) \mathbf{e}_y + \left(\delta n_z^{(1)}(t) + \delta^2 n_z^{(2)}(t)\right) \mathbf{e}_z, \tag{5a}$$

$$\mathbf{m} = -\delta^2 m_x^{(2)}(t) \mathbf{e}_x + \left(\delta m_y^{(1)}(t) + \delta^2 m_y^{(2)}(t)\right) \mathbf{e}_y + \left(\delta m_z^{(1)}(t) + \delta^2 m_z^{(2)}(t)\right) \mathbf{e}_z, \tag{5b}$$

where the second order longitudinal corrections $\delta^2 n_x^{(2)}(t)$ and $\delta^2 m_x^{(2)}(t)$ must obey

$$\delta^2 n_x^{(2)}(t) = \frac{1}{2}\left[\left(\delta m_y^{(1)}(t)\right)^2 + \left(\delta m_z^{(1)}(t)\right)^2 + \left(\delta n_y^{(1)}(t)\right)^2 + \left(\delta n_z^{(1)}(t)\right)^2\right], \tag{6a}$$

$$\delta^2 m_x^{(2)}(t) = \delta m_y^{(1)}(t) \delta n_y^{(1)}(t) + \delta m_z^{(1)}(t) \delta n_z^{(1)}(t). \tag{6b}$$

We can now insert the expansion of Eq. (5) into the expression for the spin-pumping current in Eq. (1). For the spin current component polarized along the easy axis $\mathbf{e}_x$, we

find that the leading order corrections are of second order in $\delta$. On the other hand, for the spin current components that are polarized transverse to the easy axis, the leading order corrections are first order in $\delta$. Furthermore, all leading order terms only depend on $\delta n_y^{(1)}(t)$, $\delta n_z^{(1)}(t)$, $\delta m_y^{(1)}(t)$ and $\delta m_z^{(1)}(t)$, which simplifies the following discussions considerably.

To leading order, it is then sufficient to only consider the linear corrections in $\mathbf{n}$ and $\mathbf{m}$. In the following analysis, we therefore only use the linear response expansion of the fields that are driven at the frequency $\omega_{\text{AC}}$ of the AC magnetic field

$$\mathbf{n} = \mathbf{n}_0 + \frac{1}{2}\left(\delta\mathbf{n} e^{i\omega_{\text{AC}}t} + \delta\mathbf{n}^* e^{-i\omega_{\text{AC}}t}\right), \tag{7a}$$

$$\mathbf{m} = \frac{1}{2}\left(\delta\mathbf{m} e^{i\omega_{\text{AC}}t} + \delta\mathbf{m}^* e^{-i\omega_{\text{AC}}t}\right). \tag{7b}$$

The out-of-equilibrium deviations $\delta\mathbf{n} = \delta n_y \mathbf{e}_y + \delta n_z \mathbf{e}_z$ and $\delta\mathbf{m} = \delta m_y \mathbf{e}_y + \delta m_z \mathbf{e}_z$ are perpendicular to $\mathbf{n}_0 = \mathbf{e}_x$. In general, these deviations depend on the AC field frequency and the free energy. Their magnitudes are significant only close to the resonance frequencies.

Next, we will consider the contributions to the spin backflow current of Eq. (2b). We note that the backflow current results from the primary source, the pumped spin current of Eq. (1). Therefore, the backflow current cannot exceed the leading order in the pumped spin current. In turn, this implies that the spin accumulation $\boldsymbol{\mu}_s^N(y,\omega)$ component along the $x$-direction is of a second order in $\delta$. On the other hand, the leading contribution to the spin accumulation is of a first order in the spin deviations along the transverse $y$- and $z$-directions.

To proceed, we expand the pumped spin current in a Fourier series

$$\mathbf{I}_s^p = \sum_n \mathbf{I}_n^p e^{in\omega_{\text{AC}}t}. \tag{8}$$

In this series, we decompose the spin-current into a DC term, first AC harmonics, and higher AC harmonics. The pumped DC spin current is of a second order in the deviations from the equilibrium spin configuration and it is polarized along the easy axis:

$$\mathbf{I}_0^p = \frac{i\hbar\omega_{\text{AC}} g_\perp}{4\pi}\left(\delta n_y^* \delta n_z - \delta n_z^* \delta n_y + \delta m_y^* \delta m_z - \delta m_z^* \delta m_y\right)\mathbf{e}_x. \tag{9}$$

The polarization of the first AC harmonic pumped spin current is transverse to the easy axis,

$$\mathbf{I}_1^p = \frac{i\hbar\omega_{\text{AC}} g_\perp}{4\pi}\left(\delta n_y \mathbf{e}_z - \delta n_z \mathbf{e}_y\right) \tag{10}$$

4

and $\mathbf{I}_{-1}^p = (\mathbf{I}_1^p)^*$. To the second order in the spin deviations, the higher harmonics vanish, $\mathbf{I}_n^p = 0$ when $(|n| \geq 2)$.

In the expression for the spin backflow current of Eq. (2b), we can disregard the dependence on $\mathbf{m}$ since $\mathbf{m}(\boldsymbol{\mu}_s^N \cdot \mathbf{m})$ is a third order correction. By including terms only up to second order in $\delta \mathbf{m}$ and $\delta \mathbf{n}$, we can then approximate the spin backflow current of Eq. (2b) as

$$\mathbf{I}_s^b \approx \frac{g_\perp}{2\pi} \left[ \mathbf{e}_x \left( \boldsymbol{\mu}_s^N \cdot \mathbf{n} \right) - \boldsymbol{\mu}_s^N \right]. \tag{11}$$

Similar to the spin current, we also Fourier transform the spin accumulation at the interface ($y = 0$) into DC and AC components:

$$\boldsymbol{\mu}_s^N(t) = \sum_n \boldsymbol{\mu}_n e^{in\omega_{\mathrm{AC}} t}. \tag{12}$$

The spin backflow current can also be expanded as

$$\mathbf{I}_s^b = \sum_n \mathbf{I}_n^b e^{in\omega_{\mathrm{AC}} t}, \tag{13}$$

where the harmonic components $\mathbf{I}_n^b$ are

$$\mathbf{I}_n^b = \frac{g_\perp}{2\pi} \begin{pmatrix} \frac{1}{2} \left[ \mu_{n-1}^y \delta n_y + \mu_{n-1}^z \delta n_z + \mu_{n+1}^y \delta n_y^* + \mu_{n+1}^z \delta n_z^* \right] \\ -\mu_n^y \\ -\mu_n^z \end{pmatrix}. \tag{14}$$

Note that $(\mathbf{I}_n^b)^* = \mathbf{I}_{-n}^b$, $\boldsymbol{\mu}_n^* = \boldsymbol{\mu}_{-n}$. The product $\mu_{n-1}^{y/z} \delta n_{y/z}$ ($\mu_{n+1}^{y/z} \delta n_{y/z}^*$) is an $n$-th harmonic contribution since $\delta n_{y/z}$ ($\delta n_{y/z}^*$) contains the first harmonic factor factor $e^{i\omega_{\mathrm{AC}} t}$ ($e^{-i\omega_{\mathrm{AC}} t}$).

We find a closed set of equations for the spin accumulation in the following way. We invert the relation of Eq. (4) to find the total spin current in terms of the spin accumulation. In this inversion process, it is useful to introduce the functions

$$\Gamma_1(y,\omega) = \frac{1}{2} \hbar \nu A D_N \kappa_1(\omega) \frac{\sinh\left[\kappa_1(\omega)(d_N - y)\right]}{\cosh\left[\kappa_1(\omega) d_N\right]}, \tag{15a}$$

$$\Gamma_2(y,\omega) = \frac{\hbar \nu A D_N}{4} \left[ \kappa_2(\omega) \frac{\sinh\left[\kappa_2(\omega)(d_N - y)\right]}{\cosh\left[\kappa_2(\omega) d_N\right]} + \kappa_3(\omega) \frac{\sinh\left[\kappa_3(\omega)(d_N - y)\right]}{\cosh\left[\kappa_3(\omega) d_N\right]} \right], \tag{15b}$$

$$\Gamma_3(y,\omega) = \frac{i\hbar \nu A D_N}{4} \left[ \kappa_2(\omega) \frac{\sinh\left[\kappa_2(\omega)(d_N - y)\right]}{\cosh\left[\kappa_2(\omega) d_N\right]} - \kappa_3(\omega) \frac{\sinh\left[\kappa_3(\omega)(d_N - y)\right]}{\cosh\left[\kappa_3(\omega) d_N\right]} \right], \tag{15c}$$

and $\Gamma_i(\omega) = \Gamma_i(y = 0, \omega)$ ($i = 1, 2, 3$). The resulting expression for the total spin current should equal the sum of the pumped spin current of Eq. (9) and Eq. (10), and the backflow

spin current of Eq. (14). Moreover, the expressions must hold for each component of the Fourier expansion since they should be valid at all times. Then, we find that the spin accumulation along the easy axis at the interface is:

$$\mu_n^x = \frac{1}{\Gamma_1(n\omega_{\text{AC}})} \left[ I_n^{p(x)} + \frac{g_\perp}{4\pi} \left( \mu_{n-1}^y \delta n_y + \mu_{n-1}^z \delta n_z + \mu_{n+1}^y \delta n_y^* + \mu_{n+1}^z \delta n_z^* \right) \right]. \qquad (16)$$

We see that the components along the easy axis are coupled to the perpendicular components in the $yz$-plane due to the backflow current in Eq. (14). The magnetic field aligned along the easy axis couples the transverse components of the spin accumulation in the diffusion equation (3). Hence, to find the perpendicular components of the spin accumulation at the interface, we have to solve the matrix equation for the transverse components

$$\begin{pmatrix} \Gamma_2(n\omega_{\text{AC}}) + \frac{g_\perp}{2\pi} & \Gamma_3(n\omega_{\text{AC}}) \\ -\Gamma_3(n\omega_{\text{AC}}) & \Gamma_2(n\omega_{\text{AC}}) + \frac{g_\perp}{2\pi} \end{pmatrix} \begin{pmatrix} \mu_n^y \\ \mu_n^z \end{pmatrix} = \begin{pmatrix} I_n^{p(y)} \\ I_n^{p(z)} \end{pmatrix} = \mathbf{I}_n(y=0, n\omega_{\text{AC}}) - \mathbf{I}_n^b(y=0, n\omega_{\text{AC}}). \qquad (17)$$

In the absence of a magnetic field $\Gamma_3(\omega) \to 0$, the coupling between the transverse components $\mu_n^y$ and $\mu_n^z$ vanishes. We solve Eq. (17) for $\mu_n^{y,z}$ and find that

$$\begin{pmatrix} \mu_n^y \\ \mu_n^z \end{pmatrix} = \frac{1}{\left(\Gamma_2(n\omega_{\text{AC}}) + \frac{g_\perp}{2\pi}\right)^2 + \Gamma_3^2(n\omega_{\text{AC}})} \begin{pmatrix} \Gamma_2(n\omega_{\text{AC}}) + \frac{g_\perp}{2\pi} & -\Gamma_3(n\omega_{\text{AC}}) \\ \Gamma_3(n\omega_{\text{AC}}) & \Gamma_2(n\omega_{\text{AC}}) + \frac{g_\perp}{2\pi} \end{pmatrix} \begin{pmatrix} I_n^{p(y)} \\ I_n^{p(z)} \end{pmatrix}. \qquad (18)$$

Since $\left(\Gamma_2(n\omega_{\text{AC}}) + \frac{g_\perp}{2\pi}\right)^2 + \Gamma_3^2(n\omega_{\text{AC}})$ is finite for the parameters of interest and $I_n^{p(y,z)} = 0$ for $|n| \neq 1$, we find that $\mu_n^{y,z} = 0$ for $|n| \neq 1$. We have then determined the spin accumulation at the interface in the normal metal along the transverse $y$ and $z$ directions:

$$\mu_1^y = -\frac{i\hbar\omega_{\text{AC}} g_\perp}{4\pi} \frac{\left(\Gamma_2(\omega_{\text{AC}}) + \frac{g_\perp}{2\pi}\right)\delta n_z + \Gamma_3(\omega_{\text{AC}})\delta n_y}{\left(\Gamma_2(\omega_{\text{AC}}) + \frac{g_\perp}{2\pi}\right)^2 + \Gamma_3^2(\omega_{\text{AC}})}, \qquad (19a)$$

$$\mu_1^z = \frac{i\hbar\omega_{\text{AC}} g_\perp}{4\pi} \frac{\left(\Gamma_2(\omega_{\text{AC}}) + \frac{g_\perp}{2\pi}\right)\delta n_y - \Gamma_3(\omega_{\text{AC}})\delta n_z}{\left(\Gamma_2(\omega_{\text{AC}}) + \frac{g_\perp}{2\pi}\right)^2 + \Gamma_3^2(\omega_{\text{AC}})}. \qquad (19b)$$

These transverse solutions can then be used to obtain the solutions for the longitudinal component, $\mu_n^x$, from Eq. (16). The pumping component $I_n^{p(x)}$ is only non-zero when $n = 0$, thereby only contributing to the DC component $\mu_0^x$. However, there are also second harmonics in the longitudinal component of the spin accumulation. This is caused by the coupling with $\mu_{\pm 1}^{y,z}$ from the backflow, the only finite components of $\mu_n^{y,z}$. In summary, we find that $\mu_n^x$ is finite when $n = 0, \pm 2$, and zero for all other values of $n$.

The Fourier components of the total spin current at the interface ($y = 0$) then become

$$\mathbf{I}_0 = \Gamma_1(0)\mu_0^x \mathbf{e}_x, \tag{20a}$$

$$\mathbf{I}_1 = [\Gamma_2(\omega_{\text{AC}})\mu_1^y + \Gamma_3(\omega_{\text{AC}})\mu_1^z]\mathbf{e}_y + [\Gamma_2(\omega_{\text{AC}})\mu_1^z - \Gamma_3(\omega_{\text{AC}})\mu_1^y]\mathbf{e}_z, \tag{20b}$$

$$\mathbf{I}_2 = \Gamma_1(2\omega_{\text{AC}})\mu_2^x \mathbf{e}_x \tag{20c}$$

and all higher harmonics vanish, $\mathbf{I}_n = 0 \quad (|n| \geq 3)$. Also, $\mathbf{I}_{-n} = (\mathbf{I}_n)^*$. The total spatiotemporal spin current in the normal metal is then

$$\mathbf{I}_s^N(y,t) = \left[\Gamma_1(y,0)\mu_0^x + 2\text{Re}\left(\Gamma_1(y,2\omega_{\text{AC}})\mu_2^x e^{2i\omega_{\text{AC}}t}\right)\right]\mathbf{e}_x$$
$$+ 2\text{Re}\left[\Gamma_2(y,\omega_{\text{AC}})\mu_1^y e^{i\omega_{\text{AC}}t} + \Gamma_3(y,\omega_{\text{AC}})\mu_1^z e^{i\omega_{\text{AC}}t}\right]\mathbf{e}_y$$
$$+ 2\text{Re}\left[\Gamma_2(y,\omega_{\text{AC}})\mu_1^z e^{i\omega_{\text{AC}}t} - \Gamma_3(y,\omega_{\text{AC}})\mu_1^y e^{i\omega_{\text{AC}}t}\right]\mathbf{e}_z. \tag{21}$$

We will now use this result to compute the inverse spin Hall effect (ISHE) to the lowest order in the spin-Hall angle. The charge current in the normal metal generated by the ISHE is[6,7]

$$\mathbf{j}_c^{\text{ISHE}}(y,t) = \theta_N \frac{2e}{A\hbar}\mathbf{e}_y \times \mathbf{I}_s^N(y,t), \tag{22}$$

where $\theta_N$ is the spin Hall angle in the normal metal. This charge current causes a build-up of charge accumulation at the interface. In turn, the charge accumulation generates a counter diffusion charge flow so that the net charge current in the open system vanishes. This electric field is then

$$\mathbf{E}(t) = -\frac{2\theta_N e}{A\hbar\sigma_N d_N}\mathbf{e}_y \times \int_0^{d_N} \mathbf{I}_s^N(y,t)\text{d}y = E_x^{\text{AC}}(t)\mathbf{e}_x + \left(E_z^{\text{DC}} + E_z^{\text{AC}}(t)\right)\mathbf{e}_z, \tag{23}$$

where $\sigma_N$ is the conductivity of the normal metal. From this we find that the DC electric field becomes

$$E_z^{\text{DC}} = \frac{\theta_N e\nu D_N}{\sigma_N d_N}\left(1 - \frac{1}{\cosh(d_N/\lambda_{\text{sd}}^N)}\right)\mu_0^x. \tag{24}$$

The first harmonic AC component is

$$E_x^{\text{AC}}(t) = \frac{\theta_N e\nu D_N}{\sigma_N d_N}\text{Re}\left[\left(\frac{1}{\cosh(\kappa_3(\omega_{\text{AC}})d_N)} - \frac{1}{\cosh(\kappa_2(\omega_{\text{AC}})d_N)}\right)i\mu_1^y e^{i\omega_{\text{AC}}t}\right.$$
$$\left. - \left(2 - \frac{1}{\cosh(\kappa_2(\omega_{\text{AC}})d_N)} - \frac{1}{\cosh(\kappa_3(\omega_{\text{AC}})d_N)}\right)\mu_1^z e^{i\omega_{\text{AC}}t}\right] \tag{25}$$

Finally, the second harmonic AC component is

$$E_z^{\text{AC}}(t) = \frac{2\theta_N e\nu D_N}{\sigma_N d_N}\text{Re}\left[\left(1 - \frac{1}{\cosh(\kappa_1(2\omega_{\text{AC}})d_N)}\right)\mu_2^x e^{2i\omega_{\text{AC}}t}\right]. \tag{26}$$

7

While the DC component and the second harmonic AC component are quadratic in the spin deviations, the first harmonic AC component is linear in the perturbations.

---



\end{document}